\documentclass[conference]{IEEEtran}

\usepackage[noend]{algorithmic}
\pdfoutput=1
\ifCLASSOPTIONcompsoc
  \usepackage[caption=false,font=normalsize,labelfont=sf,textfont=sf]{subfig}
\else
  \usepackage[caption=false,font=footnotesize]{subfig}
\fi
\usepackage{graphicx}
\usepackage{float,times,mathptmx}
\usepackage{amsmath,bm}
\usepackage{varwidth}
\usepackage{cite,booktabs}
\usepackage{amsfonts}

\IEEEoverridecommandlockouts

\begin{document}

\title{\resizebox{\textwidth}{!}{Generative Adversarial Learning  for Spectrum Sensing}}
\author{Kemal Davaslioglu and Yalin E. Sagduyu\\
Intelligent Automation, Inc., Rockville, MD 20855, USA\\
\{kdavaslioglu, ysagduyu\}@i-a-i.com
}

\maketitle

\begin{abstract}
A novel approach of training data augmentation and domain adaptation is presented to support machine learning applications for cognitive radio. Machine learning provides effective tools to automate cognitive radio functionalities by reliably extracting and learning intrinsic spectrum dynamics. However, there are two important challenges to overcome, in order to fully utilize the machine learning benefits with cognitive radios. First, machine learning requires significant amount of truthed data to capture complex channel and emitter characteristics, and train the underlying algorithm (e.g., a classifier). Second, the training data that has been identified for one spectrum environment cannot be used for another one (e.g., after channel and emitter conditions change). To address these challenges, a generative adversarial network (GAN) with deep learning structures is used to 1)~generate additional synthetic training data to improve classifier accuracy, and 2) adapt training data to spectrum dynamics. This approach is applied to spectrum sensing by assuming only limited training data without knowledge of spectrum statistics. Machine learning classifiers are trained with limited, augmented and adapted training data to detect signals. Results show that training data augmentation increases the classifier accuracy significantly and this increase is sustained with domain adaptation as spectrum conditions change.
\end{abstract}


\IEEEpeerreviewmaketitle

\section{Introduction}
\emph{Cognitive radio} aims to discover and utilize limited and dynamic spectrum opportunities across time, frequency and space dimensions \cite{Mitola1999}. Supported by the emergence of cost-effective software-defined radio (SDR) platforms, it becomes  feasible to perform detection, classification and prediction tasks (such as spectrum sensing and automatic modulation recognition) with cognitive radios \cite{Clancy2007, Thilina2013}. The underlying need for automatic decision making can be addressed by \emph{machine learning} that offers cognitive radios the ability to learn without being explicitly or rigidly programmed. One example is the use of standard machine learning methods such as support vector machine (SVM) for modulation classification \cite{OShea2016}. Another example is the use of convolutional neural networks for spectrum sensing \cite{Lee2017}.

However, there are two prominent challenges to overcome, in order to realize the benefits of machine learning for cognitive radio applications.
\begin{enumerate}
  \item \emph{Challenge 1}: Supervised machine learning (especially deep learning) requires significant number of training data samples (with ground truth labels) to capture spectrum dynamics with complex characteristics of channels and emitters. However, it is typically expensive (and often not even feasible) to collect the required sufficiently comprehensive and representative set of training data under all potential spectrum conditions. In particular, a long process of spectrum sensing (to collect more training data samples) may leave very limited time for data transmission in the case of dynamic spectrum access (DSA). Similarly, if a jammer is engaged in a long process of spectrum sensing, it may lose the opportunity to jam ongoing transmissions. Therefore, it is essential to limit the sensing period to a small number of training samples. This situation becomes more challenging, when multiple channels are available to sense.
  \item \emph{Challenge 2}: When the underlying spectrum environment (scenario, background, or target set) changes, a new dataset of truthed training data is needed to retrain classifiers; otherwise, classifiers trained with old data under a restricted set of spectrum conditions (such as those obtained in offline in-lab measurements) cannot be reliably used under new conditions (such as those needed for outdoor experiments) because they do not reflect the nature of the new test data.
\end{enumerate}

We address these challenges by generating \emph{synthetic training data} for machine learning though the proposed SAGA (Spectrum Augmentation/Adaptation with Generative Adversarial network) approach. SAGA leverages and advances \emph{Generative Adversarial Networks} (GANs) \cite{Goodfellow2014} supported with \emph{deep learning} and \emph{autoencoders} \cite{Hinton2006} to generate synthetic data to (re)train machine learning classifiers, when the number of training samples is insufficient or the environment represented by the existing training data changes over time. Powered by GANs, \emph{adversarial learning} \cite{Goodfellow2016} is promising to effectively learn the underling representations of complex data sources by training complex neural network architectures. The GAN formulation introduced in \cite{Goodfellow2014} consists of two competing deep neural networks:
\begin{enumerate}
  \item the \emph{generator network} maps a source of noise to the input space, and
  \item the \emph{discriminator network}  receives either a generated (synthetic) or a real data sample and distinguishes between the two samples as generated or real.
\end{enumerate}

These two neural networks play a minimax game to build a good generative model. In addition to the traditional GAN that generates only data without labels, the conditional GAN \cite{Mirza2014} extension can generate synthetic data for a given label. The GAN has been applied to build generator models for synthetic data such as images \cite{Dumoulin2016}, graphs \cite{Liu2017} and text \cite{Rajeswar2017}; however generative adversarial learning has not been applied yet to any spectrum domain problem.

In general, the state of the art in the GAN aims to generate synthetic data without any attempt yet to adapt training data labels across changing conditions. As the environment changes from the time training data is collected until new test data is received, machine learning algorithms need to be adapted through different methods such as \emph{transfer learning} or \emph{reinforcement learning}:
\begin{itemize}
  \item \emph{transfer learning} \cite{Pan2010}: most of the underlying machine learning structure is preserved and only a small part of it is updated with respect to changes in environment conditions (e.g., most layers of neural network are preserved in deep learning and only few are updated using training data under the new environment).
  \item \emph{reinforcement learning} \cite{mnih2013}: machine learning parameters (e.g., neural network weights in deep learning) are adjusted based on performance feedback in terms of a value function.
\end{itemize}

However, both of these methods require either some training data or some performance feedback, which may not be available at all or may be limited in size or may arrive only at a small rate or after a significant delay. Supported by GAN, SAGA generates realistic spectrum domain data with labels and translates them between different and unknown spectrum conditions (without assuming any training data available under changing spectrum conditions). In this context, SAGA serves the long term purpose of \emph{on-line continuous learning}, where the training set is modified continuously to match current conditions. SAGA generates representative synthetic data at low cost without need for additional data acquisition (such as additional offline measurements), such that a high fidelity machine learning algorithm can be (re)trained to maintain low error. In particular, we apply SAGA to generate training data for the \emph{spectrum sensing} problem. We assume that channel conditions and emitter characteristics are changing over time and their realizations or statistics are unknown to the spectrum sensor. Therefore, standard methods such as energy detector \cite{Axell11,Axell2012} cannot be reliably applied, since the choice of energy threshold will vary the performance significantly depending on channel and emitter characteristics.

Furthermore, a machine learning  classifier can be trained to sense the spectrum. As it is not possible to pre-train the classifier under all possible spectrum conditions (such as all possible forms of channel and emitter characteristics), SAGA makes it practical for cognitive radios to use machine learning by augmenting and adapting their training data on the fly while switching seamlessly from one spectrum condition to another one.
\begin{figure}[t!]
  \centering
\includegraphics[width=0.75\columnwidth]{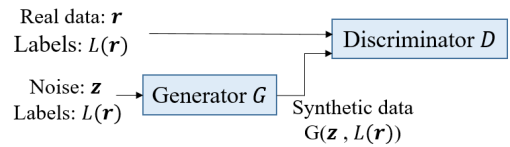}
  \caption{Conditional Generative Adversarial Network (C-GAN) that is made up of two neural networks, namely the generator ($G$) and the discriminator ($D$). Note that labels are also input to the discriminator and generator.}\label{fig:cgan}
\end{figure}

Our contribution in this paper consists of two integrated mechanisms of SAGA:
\begin{enumerate}
  \item A \emph{training data augmentation} mechanism adds synthetic data samples to the existing training dataset under a given spectrum environment. We show that adding this high-fidelity training data to retrain the classifier improves the accuracy in spectrum sensing close to the hypothetical (ideal) case when additional truthed data would be available under the same spectrum environment.
  \item A \emph{domain adaptation} mechanism generates synthetic training data for a new spectrum environment. Leveraging adversarial learning, high-fidelity synthetic training data is generated to train a new classifier for new spectrum conditions. We show that this domain adaptation approach significantly improves the accuracy in spectrum sensing under the new environment compared to the current alternative of using the classifier built for the old spectrum environment. In particular, the classifier accuracy is improved close to the hypothetical (ideal) case when training data would be available under the new spectrum environment.
\end{enumerate}

The rest of the paper is organized as follows. Section~\ref{sec:model} introduces the SAGA system model for training data augmentation and adaptation problems, and describes spectrum sensing as a machine learning classification problem. Section~\ref{sec:augmentation} presents the training data augmentation solution. Section~\ref{sec:adaptation} presents the training data adaptation solution. Section~\ref{sec:conclusion} concludes the paper.

\section{System Model}
\label{sec:model}
SAGA tackles two problems:
\begin{enumerate}
  \item \emph{Training data augmentation}: Assume some (likely little) training data $\textit{\textbf{T}}_1$, e.g., RF signals with and without emitter(s), is available for the environment $E_1$ (e.g., channel condition $h_1$). However, $\textit{\textbf{T}}_1$ does not include a sufficient number of samples to train a machine learning classifier that outputs labels $L_1$ (e.g., signal present or not) for environment $E_1$. SAGA applies conditional GAN \cite{Mirza2014} to generate more samples of $\textit{\textbf{T}}_1$ with labels $L(\textit{\textbf{T}}_1)$ that are used to train a high fidelity classifier.
  \item \emph{Domain adaptation}: Environment changes from $E_1$ (e.g., channel condition $h_1$) to $E_2$ (e.g., channel condition $h_2$). Although some training data $\textit{\textbf{T}}_1$ with truthed labels $L(\textit{\textbf{T}}_1)$ exists for $E_1$, we assume that there are no labels $L(\textit{\textbf{T}}_2)$ available that are assigned to real data samples $\textit{\textbf{T}}_2$ for $E_2$. SAGA extends the GAN formulation to generate labeled synthetic data $\textit{\textbf{T}}_2'$ with $L(\textit{\textbf{T}}_2')$ by automatically modifying samples $\textit{\textbf{T}}_1$ to resemble $\textit{\textbf{T}}_2$, while aiming to preserve the labels. This is done by a) determining the inverse of a generative model in $E_1$ (through the use of bidirectional GAN \cite{Donahue2016}), and b) passing real data in $E_1$ through an autoencoder and feeding it along with its labels to train another GAN, whose generative model $G$ is used to generate synthetic data.
\end{enumerate}

The GAN, depicted in Figure~\ref{fig:cgan}, consists of two competing deep neural networks \cite{Goodfellow2014}: the \emph{generator network} that maps a source of noise to the input space and the \emph{discriminator network} that receives either a generated (synthetic) or a real data sample and distinguishes between the two samples as generated or real. The goal is to train the generator until it learns to fool the discriminator. In theory, these models are capable of modeling an arbitrarily complex probability distribution. Mathematically, the goal is to solve the \emph{minimax} objective:
\begin{eqnarray}\label{Eqn:Minimax}
\min_{G}  \max_{D}   && ⁡\mathbb{E}_{\textit{\textbf{r}} \sim \mathbb{P}_{data}}  [\log⁡(D(\textit{\textbf{r}})) ] \nonumber \\ &&  + \mathbb{E}_{\textit{\textbf{z}} \sim \mathbb{P}_z} [\log⁡(1-D(G(\textit{\textbf{z}})))],
\end{eqnarray}
where $\textit{\textbf{z}}$ is a noise input to generator $G$ with a model distribution of $\mathbb{P}_z$ and $G(\textit{\textbf{z}})$ is the generator output. Input data $\textit{\textbf{r}} \in \textit{\textbf{T}}_1$ has distribution $\mathbb{P}_{data}$ and the discriminator $D$ distinguishes between the real and generated samples. Although the objective  in (\ref{Eqn:Minimax}) is initially proposed in \cite{Goodfellow2014}, later studies (such as \cite{Arjovsky}) showed that (\ref{Eqn:Minimax}) suffers from vanishing gradients and harder to train. Therefore, the following objectives in \cite{Arjovsky} are used for the discriminator and generator
\begin{align}
L_D & =\max_{D}  ⁡\mathbb{E}_{\textit{\textbf{r}} \sim \mathbb{P}_{data}}  [\log⁡(D(\textit{\textbf{r}})) ] + \mathbb{E}_{\textit{\textbf{z}} \sim \mathbb{P}_z} [\log⁡(1-D(G(\textit{\textbf{z}})))], \nonumber \\
L_G &= \max_{G}  \mathbb{E}_{\textit{\textbf{z}} \sim \mathbb{P}_z} [\log⁡(D(G(\textit{\textbf{z}})))],
\end{align}
The discriminator and generator are trained with backpropagation of error. In the conditional GAN formulation, $D(\textit{\textbf{r}})$ and $G(\textit{\textbf{z}})$ are replaced by $D(\textit{\textbf{r}},L(\textit{\textbf{r}}))$ and $G(\textit{\textbf{z}},L(\textit{\textbf{r}}))$, respectively, to incorporate the labels $L(\textit{\textbf{r}})$. This iterative game at the optimality minimizes the value function that minimizes the Jensen-Shannon divergence between data and model distributions.

We apply SAGA to the \emph{spectrum sensing} problem, where the goal is to detect the presence of an emitter from spectrum measurements. Although this is a standard hypothesis testing problem, we do not assume any knowledge of received signal characteristics available under a given class (or label), namely whether an emitter is present, or not. Therefore, data-driven models (instead of model-driven) methods can be effective for this classification problem. For numerical results, we assume the received signal on a channel is either noise or some signal(s) added to noise. In particular, there are two types of labels $L(\textit{\textbf{r}})$ for time series of received data $\textit{\textbf{r}}[n]=[r_1[n],\ldots,r_N[n]]$:
\begin{eqnarray}\label{HypothesisTesting}
\textit{\textbf{r}}[n] = \begin{cases} \textit{\textbf{w}}[n], & L = \{\text{no emitter}\}, \\
\textit{\textbf{h}}[n]\textit{\textbf{x}}[n] + \textit{\textbf{w}}[n], & L = \{\text{emitter}\}, \end{cases}
\end{eqnarray}
where the transmitted signal, the channel gain (with Rayleigh distribution) and additive white Gaussian noise are denoted by $\textit{\textbf{x}}[n]$, $\textit{\textbf{h}}[n]$ and $\textit{\textbf{w}}[n]$, respectively, for the $n$th sample (note that the $n$th sample of the signal, nose and channel data are represented by vectors, where each vector entry corresponds to one particular channel).
\begin{figure}[t!]
  \centering
   \includegraphics[width=\columnwidth]{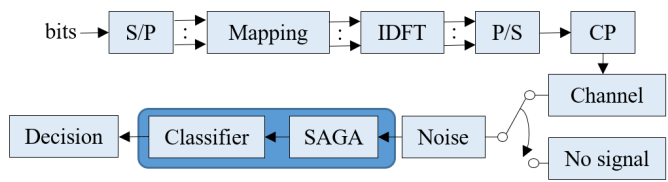}
  \caption{OFDM transmitter and receiver structure with SAGA for spectrum sensing.}
  \label{fig:ofdm}
\end{figure}

The signal of the emitter is generated from an OFDM system (depicted in Figure~\ref{fig:ofdm}), where $K$ consecutive OFDM signals out of 16QAM modulated data are transmitted. Bits are passed through a serial to parallel (S/P) converter and subcarrier mapping for data and pilot carriers. Inverse discrete Fourier transform (IDFT) of subcarriers is taken and parallel to serial conversion is carried out. Cyclic prefix is added and symbols are transmitted over the channel. The received signal is a vector of $N$~samples  in which there are $N_d$~data samples (typically the block size of the inverse fast Fourier transform, IFFT) and $N_c$ cyclic prefix (CP) samples such that $N = K(N_c+N_d)$. In general, detectors studied in the literature (e.g., energy, covariance, eigenvalue, and cyclostationary detectors) are based on the first- and second-order moments of the transmitted signal, see e.g., \cite{Axell2012,Axell11,Fitzek}. However, in most cases, the accuracy depends on the knowledge (and accuracy) of $\textit{\textbf{w}}$, $\textit{\textbf{h}}$ and $\textit{\textbf{x}}$ statistics and the knowledge of $N_c$ and $N_d$.
In particular, when the channel or noise statistics are unknown, or when there are different transmission protocols that operate in the same band, or the parameters $N_c$ and $N_d$ are dynamically selected, receivers need to be re-tuned for spectrum sensing. Instead, SAGA only observes the training data $\textit{\textbf{T}}$ and does not know any statistics or protocol information regarding $\textit{\textbf{w}}$, $\textit{\textbf{h}}$ or~$\textit{\textbf{x}}$.
\begin{figure}
  \centering
   \includegraphics[width=0.75\columnwidth]{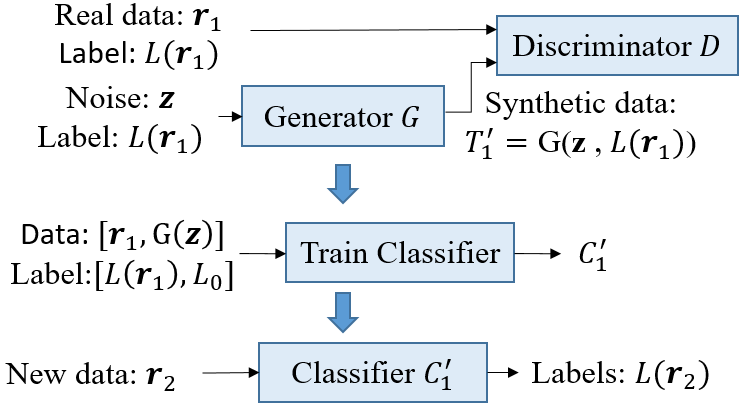}
  \caption{Workflow for training data augmentation. The first step trains a CGAN, and the second step utilizes the real samples and synthetic samples generated by the CGAN to train a classifier $C_1^{'}$. When new data samples arrive in the third step, they are labeled using the classifier trained with the augmented samples.}
  \label{fig:data_aug_workflow}
\end{figure}

\section{Spectrum Training Data Augmentation}
\label{sec:augmentation}

Figure~\ref{fig:data_aug_workflow} depicts the SAGA workflow for the \emph{training data augmentation} problem. We assume that there is some (little) training data $\textit{\textbf{T}}_1$ available in spectrum environment $E_1$ (with fixed but unknown statistics of $\textit{\textbf{h}}[n]$, $\textit{\textbf{x}}[n]$ and $\textit{\textbf{w}}[n]$). We build a spectrum sensing classifier that detects the presence of a signal using the received signal. We note that we do not assume any pre-processing of data at the receiver (e.g., removing CP or taking the Fast Fourier Transform (FFT) of the signal) except for digitizing the analog signal. Our objective for data augmentation is to generate additional synthetic training data to improve the accuracy of the spectrum sensing classifier. Thus, we employ a GAN to generate new synthetic samples $\textit{\textbf{T}}_1^{'}$ and to train the classifier $C$ with the concatenation of real data samples $\textit{\textbf{T}}_1$ and synthetic data samples $\textit{\textbf{T}}_1^{'}$.

Conditional GAN shown in Figure~\ref{fig:data_aug_workflow} is used in SAGA with two types of data samples: 
\begin{enumerate}
  \item \emph{Upper branch of the discriminator}: real training samples in $E_1$.
  \item \emph{Lower branch of the discriminator}: synthetic samples (generated by SAGA) for $E_1$.
\end{enumerate}
The SAGA workflow for training data augmentation, illustrated in Figure~\ref{fig:data_aug_workflow}, consists of the following steps:
\begin{enumerate}
\item \emph{GAN Training}: A conditional GAN is trained using real training samples. The generator learns to synthesize new data samples.
\item \emph{Classifier Training with augmented data}: The synthetic data samples are used, along with the real samples, to train a classifier.  
\item \emph{Classification}: As new data comes in, classifier $C$ is used to classify the presence of an emitter.
\end{enumerate}

For numerical results, we consider the transmission of $N_d = 32$ OFDM subcarriers with CP of length $N_c = 8$ over a unit-normal Rayleigh channel with $4$~channel taps. In the conditional GAN, we employ deep learning structures that correspond to three layers of neural network (feed-forward neural network) with $100$ neurons at each layer for both discriminator and generator. In the hidden layers, leaky rectifying linear unit (leaky-RELU) is used as activation. A leaky-RELU performs $f(x) = \max(\alpha x, x)$ operation and we use $\alpha = 0.2$. At the output layer, sigmoid activation function is used. OFDM spectrum sensing simulation is implemented in TensorFlow~\cite{Tensorflow}.

\begin{figure}[t!]
  \centering
  \includegraphics[width=0.75\columnwidth]{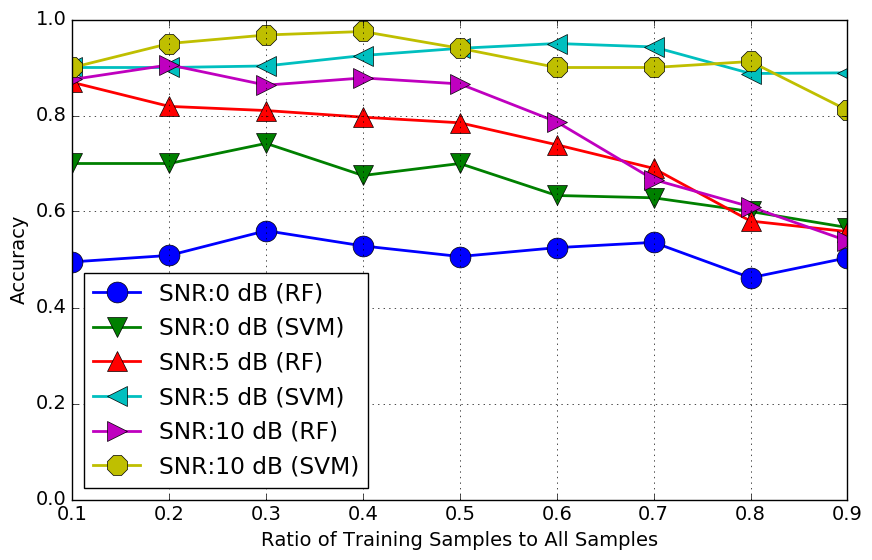}
  \caption{Accuracy of classifiers with real data samples.}\label{fig:augInitial}
\end{figure}

We evaluate the performance of two classifiers used for spectrum sensing: random forest (RF) and SVM using radial basis function (RBF) kernels. In general, RF is fast to train but may not always achieve high accuracy, whereas SVM with RBF kernel has high accuracy but requires more computational power. The accuracy of RF and SVM classifiers is evaluated at signal-to-noise ratios (SNRs) of $0$~dB, $5$~dB, and $10$~dB in environment $E_1$. Consider two cases:
\begin{enumerate}
  \item \emph{Baseline}: Classifier $C_1$ that is trained with $N_{r}$ real training data samples 
  \item \emph{Data augmentation}: Classifier $C_2$ is trained with $N_{r}$ real training data samples and $N_{s}$ samples generated by SAGA.
\end{enumerate}
In our evaluation, as an initial step, the two classifiers (one as baseline and the other one supported data augmentation) are trained with real data samples for illustration. The accuracies of both classifiers with respect to different portions allocated for training and testing are presented in Figure~\ref{fig:augInitial}. It needs to be noted that this parameter drastically changes the classifier accuracy as it determines the overfit versus underfit trade-off. The binary classification accuracy of detecting the presence of an emitter in $E_1$ is presented in Table~\ref{Table:Data_aug} at the best ratios of training samples to total samples.  SVM classifier with kernel basis outperforms RF classifier for most cases. Figure~\ref{fig:augInitial} depicts the testing accuracy as the number of training samples increase. 

\begin{table}[t!]
\centering
\caption{Accuracy of Different Training Methods}
\label{Table:Data_aug}
\begin{tabular}{cccc}\toprule Method & \multicolumn{3}{c}{Accuracy} \\ \cline{2-4}
  &  $0$~dB & $5$~dB & $10$~dB \\ \midrule
No data augmentation (RF) &  0.56 & 0.87 & 0.91\\
Data augmentation (RF) &  0.94 & 0.98 & 0.99\\
No data augmentation (SVM) &  0.74 & 0.95 & 0.97\\
Data augmentation (SVM) &  0.97 & 0.97 & 1.00\\
\bottomrule
\end{tabular}
\end{table}

\begin{figure}[t!]
\centering
\begin{tabular}{c}
\includegraphics[width=0.75\columnwidth]{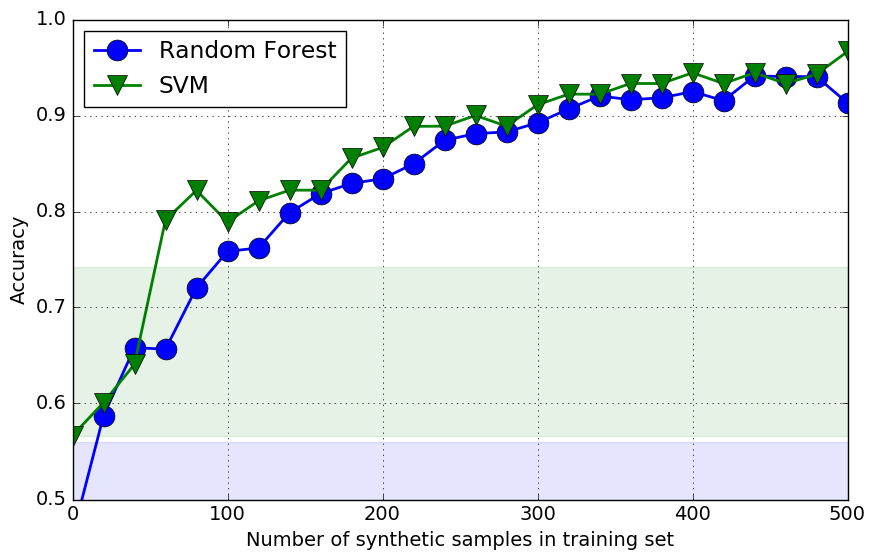} \\
\includegraphics[width=0.75\columnwidth]{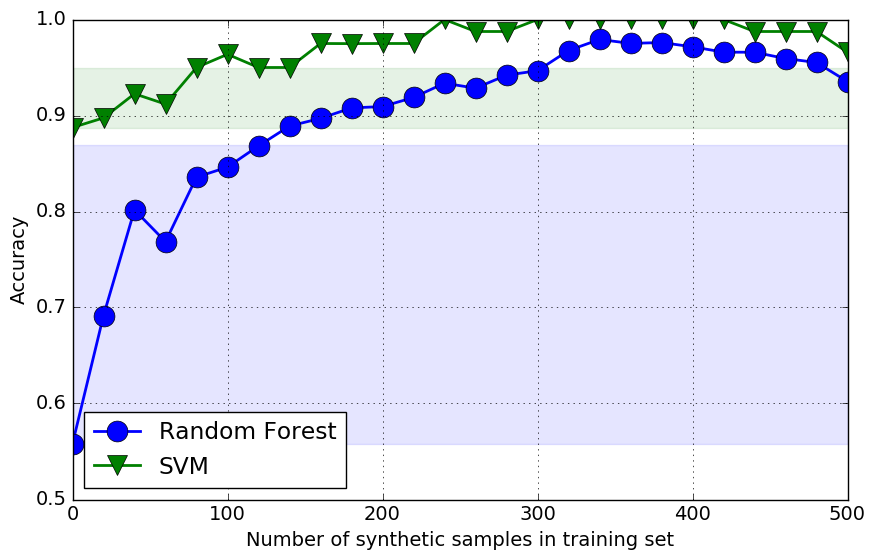} \\
\includegraphics[width=0.75\columnwidth]{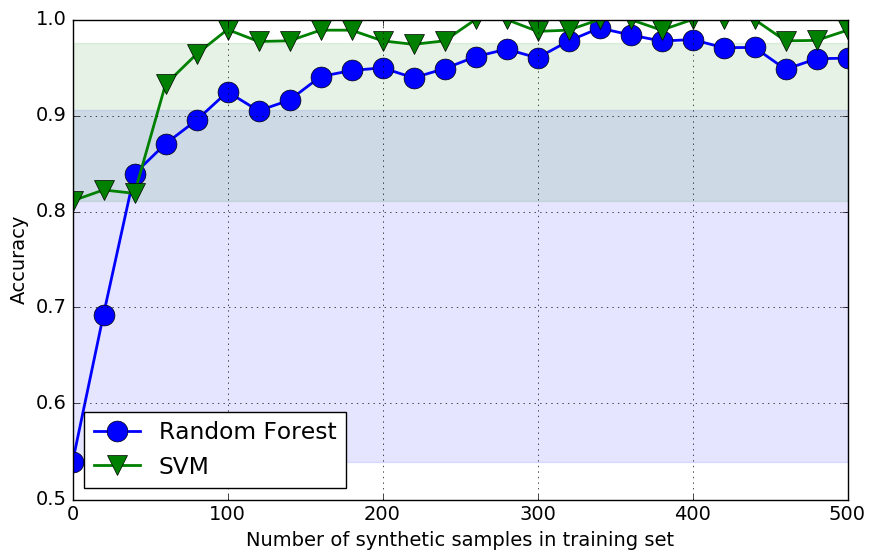}
\end{tabular}
\caption{Data augmentation results of different number of synthetic data samples for RF and SVM classifiers. Lines indicate the accuracy of classifier trained on augmented samples, colored regions denote the accuracy of regular classifier with different training data sets. From top to bottom, the figures correspond to the SNR values of $0$~dB, $5$~dB, and $10$~dB, respectively. Figure is best viewed in color.}
\label{fig:DataAugResults}
\end{figure}

For $100$ real samples, we divide the samples into training and test sets. To demonstrate the performance improvements achievable with SAGA, in the next step of our evaluation, we select the worst ratio of training to total number of samples at each SNR and for each classifier, and input the training samples to train the adversarial network of SAGA. After synthetic data samples are generated, the classifier is trained with the real data and synthetic data. Two types of classifiers are evaluated. Classifier $C_1$ is trained with real samples only, whereas Classifier $C_2$ is trained with real and synthetic samples. Both classifier types ($C_1$ and $C_2$) are evaluated with the same number of test samples for fairness.

Figures~\ref{fig:DataAugResults}(a)-(c) depict the accuracy of the RF and SVM classifiers at SNR values of $0$~dB, $5$~dB, and $10$~dB, respectively. The lines represent the accuracy of Classifiers $C_2$ for different number of synthetic samples added to the training set. The blue and green regions represent the accuracy of RF and SVM classifiers (of type $C_1$), where the upper and lower points of each region correspond to the maximum and minimum accuracies over all ratios, respectively. It is observed that after $100$~synthetic samples, the classifier trained with additional synthetic data samples always outperforms the classifier trained with real data samples. We note that for both classifiers and at each SNR, the accuracy improvement is significant. For example, RF classifier accuracy is increased from $56\%$ to $94\%$ at $0$~dB and SVM classifier accuracy is increased from $81\%$ to $99\%$ at $10$~dB. As expected, we find that the number of synthetic samples to be generated as an important factor. Empirically, we see that adding $4\times$ synthetic data, i.e., $N_s =4 N_r$, provides significant performance gains.



\section{Spectrum Training Data Adaptation}
\label{sec:adaptation}
\begin{figure}
  \centering
  \includegraphics[width=\columnwidth]{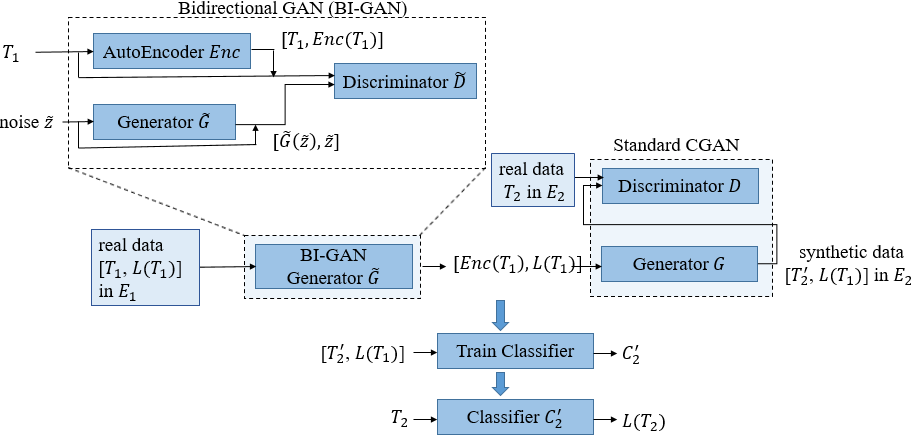}
  \caption{Workflow for training data adaptation.}\label{fig:adaptationworkflow}
\end{figure}

Suppose a classifier $C_1$ is already trained on data $\textit{\textbf{T}}_1$ with labels $L(\textit{\textbf{T}}_1)$ in $E_1$. The environment changes to $E_2$ (e.g., channel conditions or emitter characteristics) and there are no labels to train the new classifier $C_2$. The problem is to generate synthetic data that maps samples from $E_1$ to $E_2$. Figure~\ref{fig:adaptationworkflow} depicts the proposed SAGA procedure for the training data adaptation problem. We adapt bidirectional GAN framework \cite{Donahue2016}. SAGA procedure for training data adaptation consists of a bidirectional GAN, a CGAN, and a classifier. 

In Section~\ref{sec:augmentation}, we employed GAN to learn mapping random noise to realistic samples, $G(\textit{\textbf{z}})$, however, it does not give us the inverse mapping, $G^{-1}(\textit{\textbf{T}}_1)$. Bidirectional GAN helps us get this inverse mapping by using a GAN and an autoencoder that learns to take the inverse of a neural network. The samples $\textit{\textbf{T}}_1$ in $E_1$ are input to an autoencoder $\mathrm{Enc}$, while random noise is input to a generator $\tilde{G}$. The discriminator $\tilde{D}$ takes $[\textit{\textbf{T}}_1,\mathrm{Enc}(\textit{\textbf{T}}_1)]$ as real samples in the upper branch and $[\tilde{G}(\tilde{\textit{\textbf{z}}}),\tilde{\textit{\textbf{z}}}]$ as fake samples in the lower branch, and training is applied via backpropagation. When GAN training convergences, the encoder learns the inverse mapping from noise domain to the sample domain in $E_1$. Thus, we obtain the inverse mapping such that $\tilde{G}(\mathrm{Enc}(\textit{\textbf{T}}_1))= \textit{\textbf{T}}_1$. This training is done only once. 

As the environment changes from $E_1$ to $E_2$, we train a new CGAN that takes the new samples in $E_2$ as real inputs. Instead of random noise as fake inputs, we utilize the inverse mapping $\mathrm{Enc}(\textit{\textbf{T}}_1)$ of the bidirectional GAN and carry the labels in $E_1$ to train the CGAN. After CGAN training, we train a classifier $C_2'$ with domain adapted samples, $G(\mathrm{Enc}(\textit{\textbf{T}}_1))$. The classifier $C_2'$ is used to annotate the new samples in $E_2$.


\begin{figure}
  \centering
	  \includegraphics[width=0.58\columnwidth]{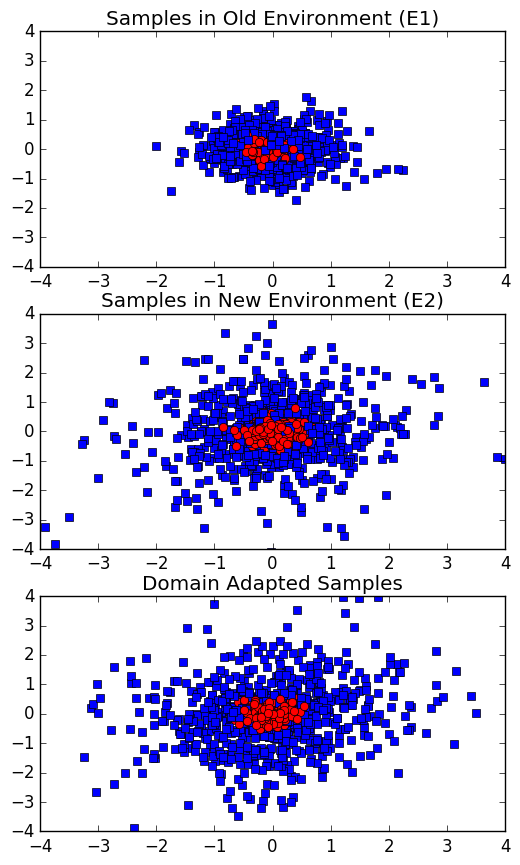}
  \caption{Distribution of data samples in different environments.}\label{fig:samplesDomains}
\end{figure}

We evaluate the performance using real samples from $E_2$. To represent two channel environments in radio frequency domain, we consider two Rayleigh fading distributions with different variances. In the first environment, we consider variance of $0.2$, whereas a variance of $2$ is considered for the second environment. The distribution of samples with and without emitters in environments $E_1$ and $E_2$ are illustrated in Figure~\ref{fig:samplesDomains}. The blue squares indicate no signal whereas red circles represent the presence of an emitter. Figure~\ref{fig:samplesDomains} depicts three types of data samples:
\begin{enumerate}
 \item upper: real training samples in $E_1$,
\item middle: real samples in $E_2$ (ideal hypothetical case), and
 \item lower: synthetic samples generated by SAGA for $E_2$.
\end{enumerate}

\begin{figure}
  \centering
  \includegraphics[width=0.9\columnwidth]{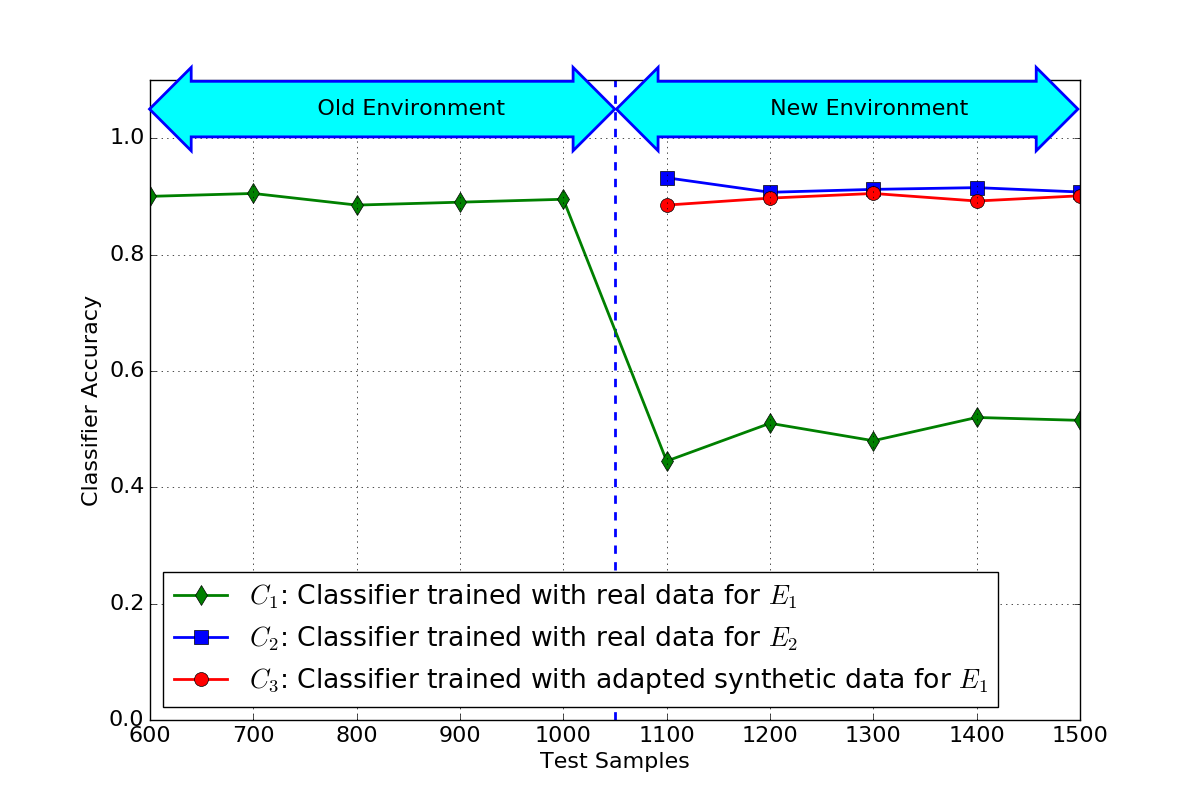}
  \caption{Classifier accuracy with and without domain adaptation.}\label{fig:adaptation}
\end{figure}

Note that the synthetic samples preserve labels in environment $E_2$. The next step is to train classifiers based on these data samples. The accuracy of SVM classifier is evaluated at $5$~dB. The classifier accuracy results (of detecting the presence of an emitter) are presented in Figure~\ref{fig:adaptation}. Three different of classifiers are considered:
\begin{enumerate}
  \item old classifier $C_1$ that is trained in $E_1$ (denoted by green),
  \item (ideal case) new classifier $C_2$ that is trained with real training data $\textit{\textbf{T}}_2$ and labels $L_2$ (denoted by blue), and
  \item new classifier $C_3$ that is trained by the data adapted by SAGA (denoted by red).
\end{enumerate}

Based on Figure~\ref{fig:adaptation}, we observe that $C_1$ performs well in $E_1$, but as the environment changes to $E_2$, its performance is severely degraded. Classifier $C_2$ constitutes the upper bound in the new environment $E_2$ as it is trained on labeled data $\textit{\textbf{T}}_2$. The classifier $C_3$ that is trained by SAGA's synthetic data performs very close to the upper bound of $C_2$.




\section{Conclusion \label{sec:conclusion}}
We presented the novel approach of SAGA that addresses two of the bottlenecks in using machine learning classifiers in cognitive radio domain, namely, the data augmentation problem and the domain adaptation problem. First, when acquiring new data is expensive or not possible, classifiers trained with the available training data may have poor performance. SAGA uses generative adversarial learning to learn and generate synthetic data samples that improve classifier accuracy though training data augmentation. The second bottleneck addressed by SAGA is domain adaptation with respect to changing spectrum environment. In this problem, labeled training data and classifiers are available for one environment, but as spectrum conditions change, no training data is available under a new environment. To address this challenge, synthetic data samples generated by SAGA can be used to train the classifier for the new environment through domain adaptation. We note that SAGA can readily be extended for wideband spectrum sensing in which a receiver listens to multiple channels to detect the presence of emitter(s) by applying training data augmentation and adaptation.

%


\begin{thebibliography}{99}
\bibitem {Mitola1999}
J. Mitola and G. Q. Maguire, ``Cognitive radio: Making software radios more personal," \emph{IEEE Personal Communications}, vol. 6, no. 4, pp. 13-18, 1999.
\bibitem {Clancy2007}
C. Clancy, H. J. Stuntebeck, and T. O'Shea, ``Applications of machine learning to cognitive radio networks," \emph{IEEE Wireless Communications},vol. 14, no. 4, pp. 47-52, 2007.
\bibitem {Thilina2013}
K. Thilina, K. W. Choi, N. Saquib, and E. Hossain, ``Machine Learning Techniques for Cooperative Spectrum Sensing in Cognitive Radio Networks," \emph{IEEE Journal on Selected Areas in Communications}, vol. 31, no. 11, pp. 2209-2221, 2013.
\bibitem{OShea2016}
T. O'Shea, J. Corgan, and C. Clancy, ``Convolutional radio modulation recognition networks," in \emph{International Conference on Engineering Applications of Neural Networks}, 2016.
\bibitem{Lee2017}
W. Lee, M. Kim, D. Cho, and R. Schober, ``Deep Sensing: Cooperative Spectrum Sensing Based on Convolutional Neural Networks," \emph{arXiv preprint arXiv:1705.08164}, 2017.
\bibitem {Goodfellow2014}
I. Goodfellow, J. Pouget-Abadie, M. Mirza, B. Xu, D. Warde-Farley, S.. Ozair, A. Courville, and Y. Bengio, ``Generative Adversarial Nets," \emph{Advances in Neural Information Processing Systems}, 2014.
\bibitem {Hinton2006}
G. Hinton and R. Salakhutdinov, ``Reducing the dimensionality of data with neural networks," \emph{Science}, vol. 313, no. 5786, pp. 504-507, July 2006.
\bibitem {Goodfellow2016}
I. Goodfellow, Y. Bengio, and A. Courville. \emph{Deep learning}.  MIT press, 2016.
\bibitem{Mirza2014}
M. Mirza and S. Osindero, ``Conditional generative adversarial nets," \emph{arXiv preprint arXiv:1411.1784} (2014).
\bibitem {Dumoulin2016}
V. Dumoulin, I. Belghazi, B. Poole, A. Lamb, M. Arjovsky, O. Mastropietro, and A. Courville, ``Adversarially learned inference," \emph{arXiv preprint arXiv:1606.00704}, 2016.
\bibitem{Liu2017}
W. Liu, P.-Y. Chen, H. Cooper, M. H. Oh, S. Yeung, and T. Suzumura, ``Can GAN Learn Topological Features of a Graph?," \emph{arXiv preprint arXiv:1707.06197}, 2017.
\bibitem{Rajeswar2017}
S. Rajeswar, S. Subramanian, F. Dutil, C. Pal, and A. Courville, ``Adversarial Generation of Natural Language," \emph{arXiv preprint arXiv:1705.10929}, 2017.
\bibitem{Pan2010}
S. Pan and Q. Yang, ``A survey on transfer learning," \emph{IEEE Transactions on Knowledge and Data Engineering}, vol. 22, no. 10, 2010.
\bibitem{mnih2013}
V. Mnih, K. Kavukcuoglu K, D. Silver, A. Graves, I. Antonoglou, D. Wierstra, and M. Riedmiller, ``Playing Atari with Deep Reinforcement Learning. \emph{arXiv preprint arXiv:1312.5602}. 2013.
\bibitem{Axell2012}
E. Axell, G. Leus, E. G. Larson, and H. V. Poor, ``Spectrum sensing for cognitive radio: State-of-the-art and recent advances," \emph{IEEE Signal Processing Magazine}, vol. 29, no. 3, pp. 101-116, May 2012.
\bibitem{Axell11} E. Axell and E. G. Larsson, ``Optimal and sub-optimal spectrum sensing of OFDM signals in known and unknown noise variance," \emph{IEEEE Journal on Selected Areas in Communications}, vol. 29, no. 2, pp. 290-304, Feb. 2011.
\bibitem{Fitzek} F. H. P. Fitzek and M. D. Katz, Cognitive Wireless Networks: Concepts, Methodologies and Visions Inspiring the Age of Enlightenment of Wireless Communications, Springer, 2007.
\bibitem{Donahue2016}
J. Donahue, P. Krahenbuhl and T. Darrell, ``Adversarial feature learning," \emph{arXiv preprint arXiv:1605.09782}. 2016.
\bibitem{Arjovsky}M. Arjovsky and L. Bottou, ``Towards principled methods for training generative adversarial networks," \emph{Proc. Int. Conf. Learning Representations}, 2017.
\bibitem{Tensorflow} M. Abadi, \emph{et al.}, ``TensorFlow: Large-scale machine learning on heterogeneous systems," 2015. Software available from tensorflow.org.

\end{thebibliography}
\end{document}